\documentclass{aa}
\usepackage{amsmath}
\usepackage{amssymb}
\usepackage{graphics}
\newcommand{\UVU}{UV$_{14}$+UVII}

\newcommand{\gccm}{\mathrm{\,g\,cm}^{-3}}
\newcommand{\sfs}{$^1\mathrm{S}_0$}
\newcommand{\sfp}{$^3\mathrm{P}_2$}
\newcommand{\comments}[1]{\bigskip\parbox[t]{0.9\linewidth}{\small{#1}}}
\newcommand{\e}{{\rm e}}
\newcommand{\df}{{\rm d}}

\begin{document}


\thesaurus{06(08.14.1; 08.05.3)} 

\title{Impact of internal heating on the thermal evolution of neutron stars}

\author{Ch. Schaab \inst{1} 
\and A. Sedrakian \inst{2}
\and F. Weber \inst{1,3}
\and M. K. Weigel \inst{4}}

\institute{Institut f{\"u}r theoretische Physik,
  Ludwig-Maximilians Universit{\"a}t, Theresienstr. 37,
  D-80333 M{\"u}nchen, Germany
\and Center for Radiophysics and Space Research,
   Cornell University, Ithaca, NY 14853, U.S.A. \\
   present address: Kernfysisch Versneller Instituut, Zernikelaan 25, NL-9747 AA
   Groningen, The Netherlands
\and Nuclear Science Division,  
  Lawrence Berkeley National Laboratory, Berkeley, CA 94720, U.S.A.
\and Sektion Physik, Ludwig-Maximilians Universit{\"a}t,
  Am Coulombwall 1, D-85748 Garching, Germany}

\offprints{Ch. Schaab, email: schaab@gsm.sue.physik.uni-muenchen.de}

\date{Received February 15, 1999 / Accepted March 4, 1999}

\authorrunning{Schaab et al.}
\titlerunning{Impact of internal heating on the thermal evolution}
\maketitle

\begin{abstract}

The impact of various competing heating processes on the thermal
evolution of neutron stars is investigated. We show that internal
heating leads to significantly enhanced surface temperatures for
pulsars of middle and old age. The heating due to thermal creep of
pinned vortices and due to outward motion of proton vortices in the
interior of the star leads to a better agreement with the observed
data in the case of enhanced cooling. The strong pinning models are
ruled out by a comparison with the cooling data on the old pulsars. For
millisecond pulsars, the heating due to thermal creep of pinned vortices
and chemical heating of the core have the largest impact on the
surface temperatures. The angular dependence of the heating rates require
two dimensional cooling simulations in general. Such a simulation is
performed for a selected case in order to check the applicability of
one-dimensional codes used in the past.

\end{abstract}

\section{Introduction} \label{sec:intro}

Simulations of the thermal evolution of neutron stars confronted with
the soft X-ray and extreme UV observations of the photon flux emitted
from their surface provide one of the most useful diagnostics of the
physical processes operating in the dense interior of such stars. The
early evolution of a neutron star is completely dominated by the
cooling via neutrino emission; at this stage the star effectively
radiates away the initial content of its thermal energy. In the later
epoch one faces a competition between cooling and heating processes,
where the heating may even dominate in the late stages of the thermal
evolution.  Eventually, they reach a thermal equilibrium phase, where
the heat generated in the star is radiated away at the same rate from
the stellar surface.

The relative importance of the various elementary processes
contributing to cooling is fairly well understood
(e.g. \cite{Richardson82}; \cite{VanRiper91};
\cite{Umeda94}; \cite{Schaab95a}; \cite{Page97a}). The
investigations performed so far have however the deficit that they did
not sufficiently clarify the relative weight of the different heating
processes. The majority of these investigations concentrated on few
preferred dissipation processes (e.g \cite{Shibazaki89};
\cite{Cheng92a}; \cite{Sedrakian93}; \cite{Reisenegger95a}) with the 
exception of the work of Van Riper (1991)\nocite{VanRiper91}, who
treats the heating in general phenomenological terms.  A meaningful
comparison of the individual processes studied in these papers is
strongly hampered by the fact that these investigations were performed
for different microphysical inputs and different levels of
sophistication concerning the cooling simulation code, preventing one
from drawing definitive conclusions about the relative importance of
the individual heating processes.

The aim of this paper is a comparative analysis of the impact of
different heating processes on the thermal evolution of neutron stars.
Our simulations take into account the most recent developments
concerning the microphysical input and employ both relativistic and
non-relativistic equations of state of superdense matter. We perform
the cooling calculations for a fully relativistic, evolutionary
simulation code.  Although internal heating is generally azimuthally
asymmetric, we shall use an one-dimensional cooling code for almost
all simulations. To demonstrate the reliability of this approximation,
a two-dimensional calculation is performed for a selected case as well
(\cite{Schaab98c}).  As far as we know, this is the first fully
two-dimensional simulation of neutron star cooling which takes
internal heating into account.

The heating processes emerge as a response to the loss of rotational
energy of the star and can be divided roughly into two groups.
Firstly, processes caused by the readjustment of the equilibrium
structure of the compact object such as a sequence of
crust/core-quakes (\cite{Cheng92a} and references therein); nuclear
reactions in cold nuclear matter with non-equilibrium composition
(picno-nuclear reactions in the crust and weak processes in the core,
see, e.g., \cite{Iida96a} and \cite{Reisenegger92a}).  Secondly,
processes related to superfluidity such as dissipative motion of the
neutron vortex lattice in the superfluid crusts (\cite{Shibazaki89}; \cite{Link91a};
\cite{Bildsten89a}; \cite{Jones90a}) and the dissipative motion of proton vortex
lattice in the superfluid core and the vorticity decay at the phase
separation boundaries, e.g. at the crust-core interface,
(\cite{Sedrakian93}; \cite{Sedrakian98a}).

The rate of dissipation associated with these processes is, in most
cases, intimately connected with the spin evolution of the star, in
particular the evolution of its magnetic moment. The time dependence
of the magnetic moment and the moment of inertia will affect the
heating rate and thus the star's cooling behaviour along with the
relation between the luminosity and the age of the star; these effects
are ignored in this study in order to constrain the actual number of
the evolutionary scenarios; we shall assume therefore a constant
magnetic moment and moment of inertia.

The paper is organised as follows. The equations of thermal
and rotational evolution of a neutron star are given in Sect.
\ref{sec:equations}.  Various heating processes associated with the
dissipative motion of vortices and the changes of the equilibrium
structure are reviewed in Sects. \ref{sec:vortex} and
\ref{sec:nonequilibrium}.  Additional physical input is discussed in
Sect. \ref{sec:inp}. The observed data are summarised in
Sect. \ref{sec:observations}. Section \ref{sec:res} presents the
results of our cooling simulations which are discussed in
Sect. \ref{sec:disc}.

\section{Thermal and rotational evolutions} \label{sec:equations}

\subsection{Equations of thermal evolution} \label{sec:eq.thermal}

The thermal evolution is governed by the coupled system of 
equations for energy balance (\cite{Thorne77}),
\begin{align}
  \frac{\partial(L\e^{2\phi})}{\partial r} &= 
    4\pi r^2 \e^\Lambda\left(-\epsilon_\nu\e^{2\phi}
      +h\e^{2\phi} -c_\mathrm{v}\frac{\partial(T\e^\phi)}{\partial t}
    \right)\, , \label{eq:ebal} \\
\intertext{and thermal energy transport,}
  \frac{\partial(T\e^{\phi})}{\partial r} &= 
    - \frac{(L\e^{2\phi})\e^{\Lambda-\phi}}{4\pi r^2\kappa} \, . \label{eq:etran}
\end{align}
This system requires as a microphysical input the neutrino emissivity
per unit volume, $\epsilon_\nu(\rho,T)$, the heat rate per unit
volume, $h(\rho,T,\Omega,\dot\Omega)$, the heat capacity per unit
volume, $c_\mathrm{v}(\rho,T)$, and the thermal conductivity,
$\kappa(\rho,T)$, where $\rho$ is the local density. The $g_{tt}$ and
$g_{rr}$ components of the Schwarzschild metric tensor $\bf g$ are
denoted by $-\e^{2\phi}$ and $\e^{2\Lambda}$, respectively, $L$ is the
luminosity, and the
other variables have their usual meaning. The boundary conditions for
(\ref{eq:ebal}) and (\ref{eq:etran}) read
\begin{align}
  L(r=0) &= 0 \, , \label{eq:bdl} \\
  T(r=r_\mathrm{m}) &= T_\mathrm{m}(r_\mathrm{m},L_\mathrm{m},M_\mathrm{m}) 
  \, , \label{eq:bdt}
\end{align} 
where $T_\mathrm{m}(r_\mathrm{m},L_\mathrm{m},M_\mathrm{m})$ is fixed
by the properties of the photosphere at the density
$\rho=\rho_\mathrm{m}=10^{10}\gccm$ (\cite{Gudmundson83}).  Since the
star's thermal evolution for times greater than, say, a few months
does not depend on the detailed initial temperature profile in the
interior of the star, one can choose, without loss of generality, an
initialy isothermal temperature distribution of $T(r)\equiv
10^{11}$~K. The set of equations for thermal evolution (\ref{eq:ebal})
and (\ref{eq:etran}) were solved numerically by means of a
Newton-Raphson type algorithm.  The microphysical input quantities
needed for the solution of these equations are given in Table
\ref{tab:ingredients} and will be discussed in detail in
Sect. \ref{sec:inp}. If one accounts for azimuthal asymmetry of the
input quantities, e.g. the heat rate $h$, one has to solve
two-dimensional thermal evolution equations, which have already been
presented in Schaab \& Weigel (1998) \nocite*{Schaab98c} along with
the numerical method for solving them.

\subsection{The normal component} \label{sec:spin.normal}

We shall start the discussion of the rotational evolution  by 
a brief review of the spin dynamics of the normal (non-superfluid)
component of the star.
The combined effect of the braking of the star's rotation via 
the emission of electromagnetic radiation, 
electron-positron wind, and gravitational waves on the spin evolution 
is commonly cast in a generic power-law form
\begin{equation} \label{eq:rot.evol}
  \dot\Omega_{\rm c}(t) = -K(t)\Omega_{\rm c}^n(t),
\end{equation}
where $\Omega_{\rm c}$ is the spin frequency of the normal component.
The functional form of $K(t)$ and $n(t)$ depends on the dominant
process assumed. If one assumes that the energy losses are due to the
magnetic dipole radiation (\cite{Goldreich71a}), then $n = 3$ and $K$
is a function of the surface magnetic field strength $B$, the angle
$\alpha$ between spin and magnetic axis, the moment of inertia of the
star $I$, and a correction factor $\gamma\equiv 1+\df\ln I/
\df\ln\Omega_{\rm c}$:
\begin{equation}
  K \propto \frac{B^2 \sin^2\alpha}{\gamma I}~.
\end{equation}
The braking index is known for the four youngest pulsars for which the
measurements of the second time derivative of the period, $P$, are
available (\cite{Lyne93a,Kaspi94a,Boyd95a}).  All four braking indices
given by $n= P\ddot P /\dot P^2$ fall into the range between 1.2 and
2.8. The deviations from the value for the magnetic dipole radiation
has been attributed to the increase of the surface magnetic field
strength (\cite{Blandford83a,Muslimov96a}), an increase of the angle
between the spin and magnetic moment axis (\cite{Beskin84a,Link97a}),
or presence of the weakly coupled superfluid in the core of the star
(\cite{Sedrakian98b}). We shall assume, in a first approach to the
problem, $K$ to be constant and $n = 3$.

With these assumptions, integration of Eq. \eqref{eq:rot.evol} yields
\begin{equation} \label{eq:rot.omega}
  \Omega_{\rm c}(t) = \left( (n-1)Kt+\Omega_{\rm i}^{-(n-1)}\right)
^{-1/(n-1)},
\end{equation}
where $\Omega_{\rm i}$ is the initial angular velocity at $t=0$. If
the initial angular velocity is large compared to the present angular
velocity, Eq. \eqref{eq:rot.omega} simplifies to
\begin{equation} 
  \Omega_{\rm c}(t) = \left( (n-1)Kt \right)^{-1/(n-1)},
\end{equation}
and the age of the pulsar can be expressed as
\begin{equation}
  t = \frac{1}{n-1}\frac{\Omega_{\rm c}(t)}{|\dot\Omega_{\rm c}(t)|}.
\end{equation}
Apart from affecting the cooling history, the time variation of $K$,
as well as a deviation of the braking index from its canonical value,
$n=3$, may lead to over- or underestimating the true age by a factor
as large as 3. A comparison of the theoretical cooling tracks with
observed data is therefore affected by this uncertainty (see
Sect. \ref{sec:observations}).

\subsection{The superfluid component} \label{sec:spin.sf}

The angular velocity of the superfluid phases adjusts to the changes in the
rotation rate of the normal component via expansion or contraction of
the neutron vortex lattice, which carries the angular momentum of the
superfluid. The mean macroscopic density of neutron vortex lines
$n_{\rm l}$ in cylindrical coordinates $(r_{\rm p},z,\phi)$ is given by
the Feynman-Onsager quantisation condition
\begin{equation}
\label{eq:vortex.dens} \kappa n_{\rm l}(r_{\rm p},t) = 2\Omega_{\rm
s}(r_{\rm p},t)+r_{\rm p} \frac{\partial\Omega_{\rm s}(r_{\rm
p},t)}{\partial r_{\rm p}}.
\end{equation}
$\Omega_{\rm s}(r_{\rm p})$ denotes the angular velocity of the
superfluid component and $\kappa = h/2m_{\rm N}$ is the quantum of
circulation carried by each vortex, where $h$ denotes the Planck
constant and $m_{\rm N}$ is the bare neutron mass.  The second term
accounts for the gradients in the distribution of vortex
lines. Because of the lack of sources and sinks for vortex lines in
the bulk of the superfluid, the temporal evolution of the vortex
number density obeys the number conservation law:
\begin{equation}\label{eq:conserv}
  \frac{\partial n_{l}}{\partial t}
      +{\vec\nabla}\cdot(n_{ l}{\vec v_{ l}}) = 0,
\end{equation}
where ${\vec v_{l}}$ is the velocity of vortex lines. Combination of
Eqs. \eqref{eq:vortex.dens} and \eqref{eq:conserv} leads to
\begin{equation} \label{eq:creeprate}
  \frac{\partial\Omega_{\rm s}}{\partial t} 
  = -\frac{1}{r_{\rm p}}\kappa n_{\rm l}v^{\rm r}_{\rm l}
  = -\left( \frac{2}{r_{\rm p}}\Omega_{\rm s} 
     +\frac{\partial\Omega_{\rm s}}{\partial r_{\rm p}} \right) 
     v^{\rm r}_{\rm l}.
\end{equation}
A decrease of the spin of the superfluid component ($\dot\Omega_{\rm
s}<0$) causes an outward radial velocity $v^{\rm r}_{\rm l}>0$. In the
following we neglect the second term in parentheses and assume
$\Omega_{\rm s}(r_{\rm p})$ to be constant throughout the superfluid
phase, since the vorticity gradients depend on the density profile of
the mutual friction coefficients, which are poorly known.
Furthermore, over the evolutionary timescale one can assume that
$\dot \Omega_c = \dot \Omega_{\rm s}$, i.e. the
deceleration rates of the superfluid and the normal components
are the same. Then the velocity of the radial motion of the vortices
(and therefore the rate of dissipation) are related to the spin-down
characteristics of the normal component via Eq. \eqref{eq:creeprate}.
In this manner the dynamics of superfluid phases is eliminated from
the system of coupled equations of secular and thermal evolution, and
the heating rates are given directly in terms the spin characteristics
of the normal component.

\section{Dissipative motion of vortex lattices} \label{sec:vortex}

We next turn to the description of the dissipation processes included
in the thermal evolution simulations. A number of dynamical coupling
regimes has been suggested in the literature for the coupling of the
superfluid component to the crusts. We shall discuss several variants
of the two main alternatives - the pinned and the corotation regimes.

\subsection{Vortex pinning at nuclei} \label{sec:pinning}

The vortex creep theories of the dynamics of neutron vortices in the
inner crust assume these vortices to be pinned to the nuclear lattice
(\cite{Anderson75a,Alpar77a,Alpar84,Epstein88a,Link91a,Pizzochero97a}).
The pinning force provides a large resistive force to the vortex
motion which is necessary to spin down the superfluid component. In
this regime, the vortices move due to the process of vortex creep,
i.e. by thermal activation from one pinning configuration to another.
The vortex creep rate has the general form (\cite{Link93})
\begin{equation} \label{eq:micro.creeprate}
  v^{\rm r}_{\rm l} = v_0 \e^{-A/T_{\rm eff}},
\end{equation}
where $v_0$ is a microscopic velocity, $A$ denotes the activation
energy for a segment of a vortex line to overcome its pinning barrier,
and $T_{\rm eff}$ is the effective temperature which includes quantum
tunnelling effects and is a function of the local temperature of the
thermal bath. These quantities depend on microscopic parameters
(e.g. the pinning energy), which themselves depend on the density, and
on the velocity difference between the superfluid and the normal
component $\delta v^\phi$.  Since the creep rate $v^{\rm r}_{\rm l}$
is given by Eq. \eqref{eq:creeprate}, Eq. \eqref{eq:micro.creeprate}
has to be inverted to obtain the velocity difference $\delta
v^\phi$. This is done by an iterative procedure similar to the one
described by Van Riper et al. (1995) \nocite{VanRiper95a}.

The rotational energy of the superfluid is converted to thermal energy
by the radial motion of the vortices. The energy dissipated by
a single vortex per unit time is $\dot E_{\rm diss}=f_{\rm M}^{\rm
r}v_{\rm l}^{\rm r}$, where $f_{\rm M}^{\rm r}=\rho_{\rm
s}\kappa\delta v^\phi$ is the radial component of the Magnus
force. The dissipated energy $h$ per unit time and volume is then
given by:
\begin{equation} \label{eq:diss.pin}
  h = \dot E_{\rm diss}n_{\rm l} = r_{\rm p}\rho_{\rm s}\delta v^\phi
  |\dot\Omega_{\rm s}|,
\end{equation}
where Eq. \eqref{eq:creeprate} has been used. The integration over the
volume of the superfluid component gives:
\begin{equation} \label{eq:pinn.heatrate}
  \int h\df V = 
  |\dot\Omega_{\rm s}| \int \rho_{\rm s}r_{\rm p}\delta v^\phi \df V.
\end{equation}
This result guarantees conservation of both energy and angular
momentum. For the implementation of $h$ in our one-dimensional cooling
code, $h$ has to be averaged over spherical shells:
\begin{equation} 
  \bar h 
  =\frac{\int h \sin\theta \df\theta\df\phi}{4\pi}
  = \frac{1}{4}\pi r\rho_{\rm s}\delta v^\phi |\dot\Omega_{\rm s}|,
\end{equation}
where we assumed that $\delta v^\phi$ is approximately constant on
spherical shells. This approximation seems to be reliable because
$\delta v^\phi$ depends only logarithmically on $r_{\rm p}$
(\cite{VanRiper95a}).

We shall consider two models for pinning: the first model (EB) uses
the parameters from Epstein \& Baym (1988) \nocite{Epstein88a} and
Link \& Epstein (1991) \nocite{Link91a}, where the pinning energy is
derived from the Ginzburg-Landau theory. In the second model (PVB) the
parameters are derived by Pizzochero et al. (1997)
\nocite{Pizzochero97a} from the Bardeen-Cooper-Schrieffer theory of
superconductivity. Below the density $1.5\times10^{13}\gccm$ the
parameters for the both models do not deviate considerably. However,
above densities $\gtrsim 1.5\times10^{13}\gccm$ the large deviations
in the pinning energy (the typical pinning energy is $\sim
10{\rm\,MeV}$ in the EB model and $\sim 1{\rm\,MeV}$ in the PVB model)
leads to rather different evolutionary scenarios and, therefore, we
shall treat both models seperately.

\subsection{Corotating regime in the crust} \label{sec:dragcrust}

The dynamics of the neutron vortices in the corotating regime is
governed by the force balance equation
\begin{equation}
  {\vec f}_{\rm M}+{\vec f}_{\rm d} = 0, 
\end{equation}
where the Magnus and the drag (resistive) forces are:
\begin{align}
  {\vec f}_{\rm M} &= -\rho_s {\vec\kappa}\times
({\vec v}_{\rm s}-{\vec v}_{\rm l}), \\
  {\vec f}_{\rm d} &= -\eta ({\vec v}_{\rm l}-{\vec v}_{\rm c}),
\end{align}
with the drag coefficient $\eta$. A decomposition of the force
balance equation into radial and azimuthal components gives
\begin{equation}
{\vec v}_{\rm l} = (v_{\rm s}-v_{\rm c})\sin\theta_{\rm d}
\cos\theta_{\rm d}{\vec e}_r
+(v_{\rm s}+v_{\rm c}\tan^2\theta_{\rm d})\cos^2\theta_{\rm d} 
{\vec e}_\phi,
\end{equation}
where the dissipation angle $\theta_{\rm d}$, following Bildsten \&
Epstein (1989)\nocite{Bildsten89a}, is defined as
\begin{equation}
  \tan\theta_{\rm d} =
\frac{\eta}{\rho_{\rm s}\kappa}.  
\end{equation}
The rate of the dissipation per unit volume is then given by
\begin{equation}
  h = -{\vec f}_{\rm d}\cdot({\vec v}_{\rm l}-{\vec v}_{\rm c})
n_{\rm free} = \rho_{\rm s}r_{\rm p}^2|\dot\Omega_{\rm s}|\omega,
\end{equation}
with the abbreviation
\begin{align}
  \omega &= \frac{v_{\rm s}-v_{\rm c}}{r} 
         = \frac{v^{\rm r}_{\rm l}}
           {r_{\rm p}\sin\theta_{\rm d}\cos\theta_{\rm d}}, \\
&= -\frac{\dot\Omega_{\rm s}}{2\Omega_{\rm s}}\frac{1}{\Phi_{\rm free}}
   \frac{1}{\sin\theta_{\rm d}\cos\theta_{\rm d}}.
\end{align}
Here, $\Phi_{\rm free}$ denotes the fraction of the free (corotating)
vortices.  This quantity is largely unknown, but for stationary
situations, one may assume that definite crustal regions support
either corotation or creep regimes and the mixing is small. Hence we
shall adopt $\Phi_{\rm free}:=n_{\rm free}/n_{\rm l}=1$ for the
corotating regime.

The dominant processes in this regime result from the coupling of the
translational motion of the vortices to the electron-phonon system
(\cite{Jones90a}). The leading order coupling implies a resistive force
constant
\begin{equation}\label{ETA0} 
\eta_0 \simeq \frac{3}{32\pi^{1/2}}\frac{a \,E_p^2}{M\xi_n^3c_s^3},
\end{equation}
where $a$ is the lattice constant, $E_p$ the pinning energy, $\xi_n$
the neutron coherence length, $M$ the (effective) mass of the nuclei,
and $c_s$ the phonon velocity (given by the unperturbed dispersion
relation). Using eq. (\ref{ETA0}) we find $\eta_0 \sim 10^5$ g
cm$^{-1}$ s$^{-1}$ at the density $\rho_s = 10^{13}$ g cm$^{-3}$
and, therefore,
$\tan\theta_d \sim 5 \times 10^{-6}$.  The processes corresponding to
higher orders in the vortex displacement are generally velocity
dependent. The next-to-leading order term, involving one-phonon
processes, is estimated as (\cite{Epstein92a,Jones92a})
\begin{equation}
\eta_1 \simeq \frac{E_p^2}
  {2\rho_s\omega a\xi_n^2\sqrt{2\pi\xi_n c_K v_L^3}},
\end{equation}
where $c_K$ is the kelvon velocity due to excitation of the
oscillatory degrees of freedom of the vortex lines moving with
velocity $v_L$. This process is sensitive to the actual excitation
spectrum of kelvons. The possible localisation of kelvons in a random
nuclear potential (\cite{Jones92a}) leads to a threshold in the
excitation spectrum beyond which a continuum of states becomes
available for excitations. In addition one requires knowledge of the
scattering amplitudes beyond the Born approximation. For typical
parameters one finds $\eta_1 \sim 10^{10}$ g cm$^{-1}$ s$^{-1}$ at a
density of $\rho_s\sim 10^{13}$ g cm$^{-3}$, and, therefore,
$\tan\theta_d
\sim 1$. However, in the stationary situations, the vortex velocities
are rather small, and we shall assume that the
leading term $\eta_0$ is the dominant one in the corotating regime.

\subsection{Electron--vortex scattering in superfluid core} 
\label{sec:electronscatt}

An estimate of the heating processes in the superfluid core requires
knowledge of vortex lattice structure that nucleates after the
superfluid phase transition. We shall pursue the point of view that
the core's magnetic field nucleates in a proton vortex lattice.  The
frictional forces in a neutron star's core are dominated by the
scattering of the quasiparticle excitations of the electron normal
Fermi-liquid by the proton vortex lattice. The reason is that the
number of proton vortices per neutron vortex, given by $n\simeq
(\sqrt{3}/2) (B/\Phi_0) d_n^2$, is of the order of $n\ge 10^{12}$ for
$B\sim 10^{12}$ G and a neutron vortex lattice constant of $d_n \sim
10^{-3}$ cm. The dominant dissipative scattering results from the
scattering of electron excitations off the quasiparticle excitations
confined in the core of proton vortex lines.  The scattering rate is a
sensitive function of the form-factor of the vortex line core and the
actual distribution of vortex lines (e.g. formation of vortex
clusters, \cite{Sedrakian98a}).  An order-of-magnitude estimate of the
dissipation rate is given by (\cite{Sedrakian93})
\begin{equation}
  h = \frac{45}{8\pi} f T
\left(\frac{\dot\Omega}{\Omega}\right)^2R^2\sin^2\theta
\cos^2\theta,
\end{equation}
where $T$ is the temperature, $R$ the radius of the superfluid core
and $f$ is defined as\footnote{Note that the multiplication sign in
the second line of Eq. (11) in Sedrakian \& Sedrakian
(1993)\nocite{Sedrakian93} should be replaced by a division sign. This
misprint is corrected here.}.
\begin{multline}
  f = 41\pi^2\alpha^2\frac{m_{\rm p}^2k_{\rm B}}{\hbar^3c}
\frac{k_{\rm F}}{k_{\rm FT}}\frac{\Delta_{\rm p}E_{\rm p}}{\hbar}
|k|\left(\frac{\xi_{\rm p}}{\lambda_{\rm p}}\right)^{2+2/3|k|} \\
\times\left(\frac{2m_{\rm p}c^2}{E_{\rm p}}\right)^{1/2}
\exp\left(-\frac{0.78\Delta_{\rm p}^2}{E_{\rm p}k_{\rm B}T}\right),
\end{multline}
where $k_{\rm F} = (3\pi^2n_{\rm p})^{1/3}$ denotes the proton Fermi
momentum, $k_{\rm FT} = \left(4k_{\rm F}m_{\rm
p}e^2/\pi\hbar^2\right)^{1/2}$ the Thomas-Fermi screening length,
$E_{\rm p} = \hbar^2k_{\rm F}^2/ {2m_{\rm p}^\ast}$ the proton
quasiparticle energy with the proton effective mass $m_{\rm p}^\ast$,
$ k \equiv (m_{\rm p}^\ast-m_{\rm p})/{m_{\rm p}}$ the entrainment
coefficient, and
\begin{align}
  \xi_{\rm p} &= \frac{\hbar^2k_{\rm F}}{\pi m_{\rm p}\Delta_{\rm p}}, \\
  \lambda_{\rm p} &= (1+|k|)^{1/2}
\left(\frac{m_{\rm p}c^2}{4\pi n_{\rm p}e^2}\right)^{1/2}
\end{align}
the proton coherence length and the magnetic field penetration length,
respectively. It should be noted that the dynamical coupling of the
proton vortex clusters to the electron liquid is controlled by the
electron flux-scattering processes, whose rate is larger than that for
the electron scattering off the vortex core quasiparticles.  While in
the latter case the scattering is inelastic (i.e.  energy is
transfered to the proton quasiparticle localised in the vortex core),
the former process of electron-flux scattering is predominantly
elastic i.e.  involves only a change of direction of the electron
momentum (but no changes in the energy).  In the classical (or
quasi-classical) limit this process corresponds to the bending of the
electron trajectory in the magnetic field of a vortex under the action
of the Lorentz force and is manifestly non-dissipative. In the quantum
limit the processes of the soft-photon emission can not be dismissed in
general; however the corresponding rates should be substantially less
than the elastric scattering rates.

\subsection{Decay of vortices} \label{sec:vortexdecay}

The proton vortices, which are dragged along by the neutron vortex
motion in the radial outward direction to the crust--core boundary,
would decay as they merge in the crust, thus releasing their
self-energy. The relevant crust-core boundary for this process should
be identified with the phase transition point between the phase with
protons bound in heavy nuclei and the phase in which they are in
continuum states. Since the crustal matter can not support a proton
supercurrent, the circulation currents of proton vortices would decay
on the ohmic dissipation timescale, which is much smaller than a
star's spin-down time scale.  The resulting heating rate at the
crust-core boundary is given by
\begin{equation}
  \int h\df V = g \frac{|\dot\Omega|}{\Omega}R^3,
\end{equation}
with
\begin{multline}
  g = \frac{1}{3}
    \left(1-\left(2\ln\frac{\lambda_{\rm p}}{\xi_{\rm p}}\right)^{-
1}\right)|k|
    \left(\frac{\xi_{\rm p}}{\lambda_{\rm p}}\right)^{2/3|k|} \\
    \times\left(\frac{\Phi_0}{4\pi\lambda_{\rm p}}\right)^2
    \ln\frac{\lambda_{\rm p}}{\xi_{\rm p}} \, .
\end{multline}
Here, $\Phi_0 = {hc}/{2e}$ is the flux quantum. The quantity $g$ is very
sensitive to variations of the  microphysical parameters such as the
proton density, the effective mass, and the gap energy at the location
of the phase transition point. We use a fixed upper limit 
for this 
process corresponding to the value of $g= 10^{28}$.

\section{Readjustment to equilibrium structure}
\label{sec:nonequilibrium}

\begin{figure} 
\resizebox{\hsize}{!}{\rotatebox{-90}{\includegraphics{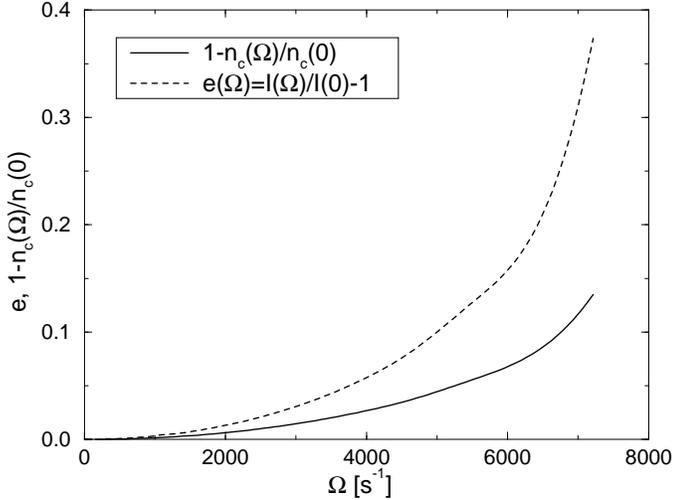}}}
\caption{Oblateness $e$ and the variation of the central density 
  $1-n_{\rm c}(\Omega)/n_{\rm c}(0)$ of rotating neutron stars versus
  angular velocity. The parameters are: $M_{\rm B}=1.58M_{\odot}$ and
  \UVU -EOS.}  \label{fig:rot}
\end{figure}

As a neutron star spins down its global structure readjusts itself
toward a more spherically symmetric configuration to minimise the sum of 
gravitational and rotational energy.  Two models of internal heating
discussed here are based on the structural readjustment of the
oblateness of the star, defined as
\begin{equation}
  e_{\rm eq} := \frac{I(\Omega)}{I(0)}-1,
\end{equation}
and of the central baryon density $n_{\rm c}$. $I(\Omega)$ denotes the
star's moment of inertia. The equilibrium values of both quantities
are determined as functions of the angular velocity by computing a
sequence of rotating neutron star models with constant baryon number
(\cite{Schaab97a}). These calculations are done by solving the general
relativistic stellar structure equations in two dimensions, in a
similar manner as in Komatsu et al. (1989)\nocite{Komatsu89a} and
Bonazzola et al. (1993)\nocite{Bonazzola93a}. As an example the
oblateness $e$ and the fraction $1-n_{\rm c}(\Omega)/n_{\rm c}(0)$
with the central density $n_{\rm c}(\Omega)$ are plotted for a neutron
star model with constant baryon mass $M_{\rm B}=1.58M_\odot$ in
Fig. \ref{fig:rot}. The underlying equation of state is the
non-relativistic \UVU -model (see Sect.
\ref{sec:inp.eos}).

\subsection{Crust cracking} \label{sec:noneq.cracking}

If both the interior and the crust are liquid, the oblateness will
continuously adjusts itself to the equilibrium value. This is the case
for the temperatures $T\gtrsim 10^8$~K (\cite{Cheng92a}). However, if
the crust solidifies, the oblateness continuously depart from its
equilibrium value causing an increasing mean stress on the crust
according to (\cite{Baym71c})
\begin{equation}
  \sigma = \tilde{\mu}(e-e_{\rm eq}),
\end{equation}
where $\tilde{\mu}$ is the mean shear modulus of the crust. 
The quantity $e_{\rm eq}$ refers to the equilibrium
value of the oblateness. If the mean stress $\sigma$ reaches some
critical value
\begin{equation} \label{eq:maxshearangle}
  \sigma_{\rm c} = \tilde{\mu}\theta_{\rm c},
\end{equation}
the crust breaks and the deviation of $e$ from its equilibrium value
is reduced. For an ideal Coulomb lattice, the value of the critical
shear angle $\theta_{\rm c}$ is of the order of magnitude
$10^{-1}-10^{-2}$ (\cite{Smolukowski70a}).  The fraction of impurities
or defects in the neutron star crusts could be considerable due to,
e.g., the fast cooling below the melting temperature of the lattice.
Smolukowski (1970)\nocite{Smolukowski70a} argues that the value of
$\theta_{\rm c}$ is then reduced to roughly $10^{-4}-10^{-5}$.

During the breaking of the crust the strain energy 
\begin{equation}
  \Delta E_{\rm strain} = 2B(e-e_{\rm eq})\Delta(e-e_{\rm eq})
\end{equation}
is released (\cite{Cheng92a}), where $\Delta(e-e_{\rm eq})\ll(e-e_{\rm
eq})$ is the change in $(e-e_{\rm eq})$, and the coefficient $B$ is
given in terms of the mean shear modulus by $B=\frac{1}{2}V_{\rm
c}\tilde{\mu}$ with $V_{\rm c}$ being the volume of the crust. We
follow Cheng et al. (1992)\nocite{Cheng92a} in assuming that
$(e-e_{\rm eq})$ is approximately equal to its maximum value
$\theta_{\rm c}$ and that the time $\Delta t$ between two successive
quakes is small compared to the time scale of thermal
evolution. Hence\footnote{Our formulas differ from those of Cheng et
al. (1992)\nocite{Cheng92a} by a factor $\theta_{\rm c}/e$, which is
$\ll 1$ for slowly rotating stars.  We therefore obtain no effect on
the cooling of slowly rotating stars, in contrast to Cheng et
al. (1992)\nocite{Cheng92a} where rather large effects are claimed.}
\begin{equation} \label{eq:strainenergy}
\frac{\df E_{\rm strain}}{\df t} \approx \frac{\Delta E_{\rm strain}}{\Delta t}
  = 2B\theta_{\rm c}\frac{\Delta(e-e_{\rm eq})}{\Delta t}
  \approx -2B\theta_{\rm c}\frac{\df e_{\rm eq}}{\df\Omega_{\rm c}}
   \dot\Omega_{\rm c}.
\end{equation}
The crust strength $B$ is given by the integral (\cite{Baym71c})
\begin{equation}
  B = \int_{V_{\rm c}} b(r)\df^3r
\end{equation}
with
\begin{equation}
  b(r) = \frac{1}{25} C_{44}(r)\left(96-166\left(\frac{r}{R}\right)^2
    +\frac{777}{10}\left(\frac{r}{R}\right)^4\right)
\end{equation}
and the shear modulus of a bcc-lattice (\cite{Mott36a})
\begin{equation}
  C_{44}(r) = 0.3711 Z^2e^2n_{\rm N} \left(\frac{2}{n_{\rm N}}\right)^{-1/3}.
\end{equation}
The total strain energy release \eqref{eq:strainenergy} can thus be expressed
in terms of a local energy loss rate:
\begin{equation}
  h = 2b\theta_{\rm c}\frac{\df e_{\rm eq}}{\df\Omega_{\rm c}}
  |\dot\Omega_{\rm c}|.
\end{equation}

\subsection{Chemical heating} \label{sec:chem}

The central density of a rotating neutron star with fixed baryon
number depends on the angular velocity (see Fig. \ref{fig:rot}). As
the star spins down, the centrifugal force decreases and
correspondingly the central density increases (by up to 20~\%). This
is accompanied by a shift of the chemical equilibrium. The matter
would maintain chemical equilibrium if the relaxation timescales for
the weak processes are small compared to the timescale of rotational
evolution. However, these timescales are found to be comparable and
the actual composition departs from chemical equilibrium, which
modifies the reaction rates (\cite{Haensel92a}) and usually leads to a
net conversion of chemical energy into thermal energy
(\cite{Reisenegger92a,Iida96a}).

For simplicity we restrict ourselves to a system containing only
neutrons, protons, electrons, and muons in $\beta$-equilibrium (as is
described by the non-relativistic EOS \UVU). The equation of state of
cold, charge neutral matter, which is not necessarily in $\beta$-equilibrium
(we assume however equal chemical potentials for
electrons and muons: $\mu_{\rm e}=\mu_\mu$.), therefore depends only
on two parameters. These are chosen to be the baryon density $n$ and
the relative proton density $x=n_{\rm p}/n$. These two parameters are
constant on closed, sphere-like surfaces $\mathcal{S}(N)$ of
constant energy density $\rho$ and pressure $P$. $N$ refers to the
number of baryons enclosed by $\mathcal{S}$. We approximate the time
derivative of the baryon density on a specific surface
$\mathcal{S}(N)$ by the expression (\cite{Reisenegger92a})
\begin{equation} \label{eq:dens.var}
  \frac{\df n}{\df t} \simeq 
    \left. \frac{n}{n_{\rm c}} \right\vert_{\Omega=0}
    \frac{\df n_{\rm c}}{\df\Omega} \frac{\df\Omega}{\df t},
\end{equation}
where $n_{\rm c}$ is the central baryon density, and $\df n_{\rm c} /
\df\Omega$ is obtained from calculations of rotating neutron star
sequences (see Fig.  \ref{fig:rot}).

The equilibrium value $x_{\rm eq}$ of the proton fraction is
determined by the following condition on the energy per baryon $E$:
\begin{align}
  \delta\mu := \frac{\partial E}{\partial x} 
  &= \frac{\partial E}{\partial n_{\rm p}}\frac{\partial n_{\rm p}}{\partial x}
    +\frac{\partial E}{\partial n_{\rm n}}\frac{\partial n_{\rm n}}{\partial x}
    +\frac{\partial E}{\partial n_{\rm e}}\frac{\partial n_{\rm e}}{\partial x}
    +\frac{\partial E}{\partial n_\mu}\frac{\partial n_\mu}{\partial x} \\
  &= \mu_{\rm p}-\mu_{\rm n}+\mu_{\rm e} = 0,
\end{align}
where we used the charge neutrality of the system, $\partial n_{\rm
p}/ \partial x = \partial n_{\rm e}/ \partial x + \partial n_\mu/
\partial x$, the condition 
$\mu_{\rm e}=\mu_\mu$, and the equation $\partial n_{\rm p}/ \partial
x=-\partial n_{\rm n}/ \partial x=n$.  The time derivative of $n$ is
linked to the time derivative of $x^{\rm eq}$ by
\begin{equation}
  \frac{\df x^{\rm eq}}{\df t} \simeq \frac{\df x^{\rm eq}}{\df n}
  \left.\frac{n}{n_{\rm c}}\right|_{\Omega=0} \frac{\df n}{\df\Omega}
    \frac{\df\Omega}{\df t},
\end{equation}
where Eq. \eqref{eq:dens.var} was used.

A deviation from chemical equilibrium corresponds to a non-vanishing
value of $\delta\mu$. The non-vanishing gradient of the total chemical
potential including both internal and external contributions causes a
diffusion of particles. If the timescale for chemical relaxation is
larger than the timescale for this diffusion, the chemical relaxation
takes place in a larger regoin of the neutron star, and $\delta\mu$
has to be averaged over such regions. It seems likely that this kind
of relaxation occurs in the core of neutron stars
(\cite{Reisenegger96a}).

The time derivative of
\begin{equation}
  \delta\mu = \frac{\partial E}{\partial x} = \left.
\frac{\partial^2E}{\partial x^2}\right\vert_{x=x_{\rm eq}}
    (x-x_{\rm eq})
\end{equation}
yields the evolution equation for $\delta\mu$:
\begin{equation}
  \frac{\partial\delta\mu}{\partial t} = \left. \frac{\partial^2E}{\partial 
x^2}\right\vert_{x=x_{\rm eq}}
    \left( \frac{\Gamma}{n}-\frac{\df x^{\rm eq}}{\df n}
\left.\frac{n}{n_{\rm c}}\right|_{\Omega=0} \frac{\df n}{\df\Omega}
    \frac{\df\Omega}{\df t}\right),
\end{equation}
where $\Gamma(\delta\mu,n)=n\partial x/ \partial t$ is the difference
between the rates per unit volume of electron capture and beta
decays. This difference gives a positive value for $\delta\mu>0$, so
that chemical equilibrium tends to be restored. The second derivative
of the energy per baryon is equal to
\begin{equation}
  \left.\frac{\partial^2E}{\partial x^2}\right\vert_{x=x_{\rm eq}} 
  = 8S(n)+\frac{1}{3}\hbar c(3\pi n)^{1/3}x^{-2/3},
\end{equation}
where $S(n)$ denotes the symmetry energy.

The average energy released by these reactions into kinetic and thus
thermal energy is equal to the difference of the Fermi energies of
protons and electrons on the one side and neutrons on the other side,
i.e.:
\begin{equation}
  h = \Gamma\delta\mu\, .
\end{equation}
At the same time, thermal energy is lost by emission of thermal
neutrinos, as it is the case for matter in chemical equilibrium. The
emission rates are however modified by the deviation from chemical
equilibrium (see \cite{Reisenegger92a} for the values of $\Gamma$ and
$\epsilon_\nu$). Like the neutrino emissivity $\epsilon_\nu$, the
reaction rate $\Gamma$ is suppressed by superfluidity unless the
deviation of chemical equilibrium $\delta\mu$ exceeds the sum of the
gap energies of the participating baryons (\cite{Reisenegger96a}).

Until now we have considered only the restoration of chemical
equilibrium in the core. The situation in the crust is similar, though
more complex. Iida \& Sato (1997)\nocite{Iida96a} studied the
Lagrangian changes in pressure associated with elements of matter due
to the spin-down in the framework of the Hartle-approximation
(\cite{Hartle67a}). By considering the nuclear processes induced above
$\rho_{\rm l}=3\times 10^{13}\gccm$ by compression or decompression of
matter, they approximate the chemical energy per unit volume and time
converted to thermal energy to
\begin{equation}
  h \approx
  nqN\sin\theta|1-1.46\cos^2\theta|\frac{2\Omega|\dot\Omega|}{(6283{\rm\, s}^{-1})^2}\, ,
\end{equation}
where $nq\approx 4{\rm\, eV\, fm}^{-3}$ is the energy per unit volume
released by one non-equilibrium process, $N\approx 15(\rho-\rho_{\rm
  l})/(\rho_{\rm tr}-\rho_{\rm l})$ is the number of processes in a
unit cell, and $\rho_{\rm tr}=1.7\times 10^{14}\gccm$ is the
transition density between core and inner crust. By averaging over
spherical shells we obtain
\begin{equation}
  h \approx 3\times 10^{21} \frac{\rho-\rho_{\rm l}}{\rho_{\rm
      tr}-\rho_{\rm l}} \frac{\Omega|\dot\Omega|}{1{\rm\, s}^{-3}}
   {\rm\, erg\, cm}^{-3}{\rm\,s}^{-1} \, .
\end{equation}

\section{Additional ingredients and comparison of the efficiency of
  the heating mechanisms} \label{sec:inp}

Apart from the rates of the various heating processes discussed in the
previous sections, some further ingredients are needed to solve the
general relativistic equations of stellar structure and evolution (see
Sect. \ref{sec:eq.thermal}). These additional ingredients are
summarised in Table  \ref{tab:ingredients}. Here we shall only discuss
the equation of state of the dense interior of a neutron star, the
neutrino emissivities, the superfluidity gaps and the photosphere,
since these ingredients are the most important ones. Further, we
will compare the efficiencies of the various heating mechanism.

\begin{table*} \centering
\caption[]{Input quantities used for the cooling
    simulations \label{tab:ingredients}}
\begin{tabular}{ll}
\hline\noalign{\smallskip}
Parameter & References \\
\noalign{\smallskip}\hline\noalign{\smallskip}
  Equations of state: \\
  \quad crust        & \cite{Haensel94a},
                       \cite{Negele73} \\
  \quad core (alternatives)  & \\
  \qquad \UVU        & \cite{Wiringa88} \\
  \qquad RHF8        & \cite{Huber97a} \\
  \hline
  Superfluidity      & see Table \ref{tab:sf} \\
  \hline
  Heat capacity      & \cite{Shapiro83},
			\cite{VanRiper91} \\
  \hline
  Thermal conductivities: \\
  \quad crust        & \cite{Itoh84a}, \cite{Itoh83a},
                       \cite{Mitake84} \\
  \quad core         & \cite{Gnedin95a} \\
  \hline
  Neutrino emissivities: \\
  \quad pair-, photon-, plasma-processes & \cite{Itoh89} \\
  \quad bremsstrahlung in the crust      & \cite{Yakovlev96a}, \cite{Haensel96a} \\
  \quad bremsstrahlung in the core & \cite{Friman79}, \cite{Kaminker97a} \\
  \quad modified Urca & \cite{Friman79}, \\
                      & \cite{Yakovlev95a} \\
  \quad direct nucleon Urca & \cite{Lattimer91} \\
  \quad direct hyperon Urca & \cite{Prakash92} \\
  \quad superfluid pair breaking & \cite{Voskresenskii87a}, \\ 
                            & \cite{Schaab95b} \\
\hline
  Photosphere:  	& \cite{Potekhin96c} \\
\noalign{\smallskip}\hline
\end{tabular}
\end{table*}

\subsection{Equation of state} \label{sec:inp.eos}

For the outer and inner crust we adopt the equations of state of
Haensel \& Pichon (1994)\nocite{Haensel94a} and Negele \& Vautherin
(1973)\nocite{Negele73}. The transition density between the ionic
crust and the core of a neutron star is taken to be $\rho_{{\rm
tr}}=1.7\times 10^{14}\gccm$ (\cite{Pethick95}).  For the present
study, we choose two models of high density matter.  The first,
non-relativistic model (\UVU) is obtained by solving the Schr\"odinger
equation by means of a variational approach (\cite{Wiringa88}). The
other model (RHF8) uses relativistic Br\"uckner-Hartree-Fock results
up to 2-3 times normal nuclear density. Hyperons are included within
the relativistic Hartree-Fock approach above this density
(\cite{Huber97a}). One important difference between these equations of
state is that the non-relativistic model treat neutron star matter as
being composed of neutrons and protons only (which are in
$\beta$-equilibrium with leptons), whereas the relativistic model
accounts for all hyperon states that become populated in the cores of
neutron stars.

\subsection{Neutrino emissivity} \label{sec:inp.emis}

The neutrino emission processes can be divided into slow and enhanced
processes depending on whether one or two baryons participate in the 
reaction. Due to
the rather different phase spaces associated with both kind of
processes the emission rates differ by several orders of magnitude.

If one neglects exotic states in neutron star matter (like quark-gluon
plasma and meson condensates) the only enhanced neutrino emission
processes are the direct nucleon and hyperon Urca processes
\begin{align}
  \mathrm{n} &\rightarrow \mathrm{p} + \lambda^- + \bar{\nu}_\lambda \\
  B_1 &\rightarrow B_2 + \lambda^- + \bar{\nu}_\lambda ,
\end{align} 
respectively, where $B_{1,2}=\mathrm{n}$, p, $\Sigma^{\pm,0}$,
$\Lambda$, $\Xi^{0,-}$ denotes the baryons and
$\lambda_{1,2}=\mathrm{e}^-$, $\mu^-$ the leptons.  Because of the
$\beta$-equilibrium the inverse reaction (with $\bar\nu$ replaced by
$\nu$) occurs at the same rate as the direct one.  The emissivities of
these processes were computed by Prakash et
al. (1992)\nocite{Prakash92}.  Simultaneous conservation of energy and
momentum requires that the triangle inequality $p_{B_1}^{\rm F} <
p_{B_2}^{\rm F} + p_{\rm e}^{\rm F}$ and the two inequalities obtained
by cyclic permutation are fulfilled for the Fermi momenta $p_i^{\rm
F}$.  If the inequalities are not fulfilled the process is extremely
unlikely to occur and the corresponding emissivity can be neglected.
The availability of the various fast processes thus depends on the
partial concentrations of each baryon species. The proton fraction,
which determines the possibility of the direct nucleon Urca process,
depends crucially on the symmetry energy, which is poorly known for
high density matter. As an example of an equation of state which
yields slow cooling we study the non-relativistic equation of state
\UVU. Due to the non-monotonic behaviour of the symmetry energy at high
densities, the triangle inequality is not fulfilled for this equation
of state. The relativistic equation of state RHF8 allows for both the
direct nucleon and the direct hyperon Urca processes for star masses
$M>1.09\,M_\odot$ and $M>1.22\,M_\odot$, respectively.

During the transition to the superfluid state of neutrons and protons
the superfluid pair breaking and formation processes become important
(\cite{Flowers76,Voskresenskii87a}). We refer to Schaab et
al. (1997)\nocite{Schaab95b} for a detailed description of these
processes and their impact on the cooling of neutron stars.

\subsection{Superfluidity} \label{sec:sf}

The pairing gaps are sensitive to the underlying microscopic model for
the nucleon-nucleon interaction. Since these models are rather
uncertain, especially at high densities, the value of the maximum gap
and the density range where pairing can occur is model dependent. The
disagreement arising from the different interaction models is enlarged
by more subtle issues of the medium polarisation
(\cite{Wambach91a,Schulze96a}) or the inclusion of relativity
(\cite{Elgaroy96c}).

\begin{table*} \centering
\caption[]{Gap energies of superfluid states in neutron star
	matter. The density ranges are calculated for the RHF8
	equation of state.  \label{tab:sf}}
\begin{tabular}{lccl}
\hline
Pairing	& $\Delta_{\rm max}$ [MeV] & Density range [fm$^{-3}$] & References \\
\hline
neutron \sfs	& 1.01 & $<\,0.16$	
& \cite{Schulze96a} \\
proton \sfs	& 0.92 & 0.10--0.31 	
& \cite{Elgaroy96a}\\
neutron \sfp	& 1.45 & $>\,0.06$ 	
& \cite{Baldo98a}, CD-Bonn potential \\
lambda \sfs	& 0.24 & 0.44--0.63 	
& \cite{Balberg97b} \\
\hline
\end{tabular}
\end{table*}

At densities of the order $2\times 10^{14}\gccm$ the $S$ state
interaction becomes repulsive for neutrons, and the $^1S_0$ neutron
gap closes.  The attractive $^3P_2-^3F_2$ state interaction leads to
the pairing in this channel at about the same densities.  For the $P$
state pairing we use the recent calculation of Baldo et
al. (1998)\nocite{Baldo98a} which is based on the modern
nucleon--nucleon CD-Bonn potential (\cite{Machleidt96a}), which
provides accurate fits to the actual nucleon-nucleon scattering data
below 350~MeV.

Field theoretical descriptions of the heavy-baryon interaction based
on the one-boson-exchange model indicate that the $\Lambda-\Lambda$
interaction has an attractive component.  From the so-called
doubly-strange hypernuclei, where two hyperons are bound in a single
nucleus (\cite{Imai92a}), one can verify this field theoretical
result, since, for instance, the separation energy of the two
$\Lambda$ hyperons, $B_{\Lambda\Lambda}$, exceeds the separation
energy of a single $\Lambda$ from the same nucleus, $B_{\Lambda}$, by
more than a factor of two. As expected according to the quark model,
the binding energy of hyperons, $\Delta
B_{\Lambda\Lambda}\!\equiv\!B_{\Lambda\Lambda}-2B_\Lambda\approx4-5\;$
MeV, is somewhat less than the corresponding binding energy of
nucleons, $\Delta B_{NN}\approx6-7\;$MeV.  Due to these similarities,
$\Lambda$ hyperon pairing is expected in high density matter in
analogy with the case of nucleons. Here, we use the $\Lambda$ hyperon
\sfs\ gaps estimated by Balberg \& Barnea (1998)\nocite{Balberg97b}. The
impact of $\Lambda$ pairing on the thermal evolution of neutron stars
has recently been investigated in Schaab et
al. (1998a)\nocite{Schaab98b}.

After the onset of the superfluid state of the respective constituents
of stellar matter the neutrino emissivities, the thermal conductivity,
and the heat capacity are suppressed by factors which behave like
$\exp(-aT_{\rm c}/T)$ for $T\ll T_{\rm c}$, where $T_{\rm c}$ denotes
the critical temperature and $a$ is some constant of order unity
(\cite{Maxwell79,Levenfish94a,Gnedin95a,Yakovlev95a}). We refer to
Levenfish \& Yakovlev (1994)\nocite{Levenfish94a} for fitting formulas
of ${\mathcal R}_{\rm sf}$, that are valid over the whole temperature
range $T<T_{\rm c}$ for both types of pairing state. The suppression
factor of the \sfp\ pairing behaves exponentially only if the neutrons
pair in the nodeless state with $m=0$. In the case of $m=2$ this
factor behaves polynomially
(\cite{Anderson61,Muzikar80,Levenfish94a}). The effects of $m=2$
pairing on the cooling behaviour is discussed in Schaab et
al. (1998b)\nocite{Schaab98a} where considerable modifications were
found.  Due to the uncertainty in of the ground state quantum numbers
of the neutron condensate we shall adopt here the more conventional
case of the $m= 0$ nodeless pairing.

\subsection{Photosphere} \label{sec:inp.photo}

While the heat conduction in the interior of the star is treated
dynamically in our simulations the actual surface temperatures are
obtained by attaching an outer envelope with $\rho \le \rho_{\rm m}$
at the boundary to the 'core'.  The validity of this approximation has
been discussed in Gudmundsson et al. (1983)\nocite{Gudmundson83}. To
obtain the surface temperatures we use the envelope calculations by
Potekhin et al. (1997)\nocite{Potekhin96c} whose non-magnetic
photospheres provide a smaller gradient for low effective surface
temperatures $T_{\rm eff}\lesssim 3\times 10^5{\rm\,K}$ than the
earlier calculations of Gudmundson et al. (1983)\nocite{Gudmundson83}
and Van Riper (1988)\nocite{VanRiper88}, mainly due to a refined
treatment of the electron conductivity in the non-degenerate
region. This yields a slower cooling in the photon cooling era for
stars older than $\gtrsim10^6{\rm\,yr}$.

Even a rather small amount of accreted matter of say $\Delta
M\sim10^{-16}M_\odot$, substantially reduces the thermal insulation of
the photosphere. The effect is highest for $\Delta
M\sim10^{-7}M_\odot$, where the accreted hydrogen is partly burned
into helium and carbon. Additionally accreted matter will be converted
into iron, leaving the thermal insulation of the star practically
unchanged (\cite{Potekhin96c}). The effect of accreted envelopes on
cooling of neutron stars are discussed in Sect. \ref{sec:res.accr}.

\subsection{Comparison of the efficiency of the heating mechanism} 
\label{sec:inp.comp}

\begin{figure*} 
\centering\resizebox{0.8\hsize}{!}{\rotatebox{-90}{\includegraphics{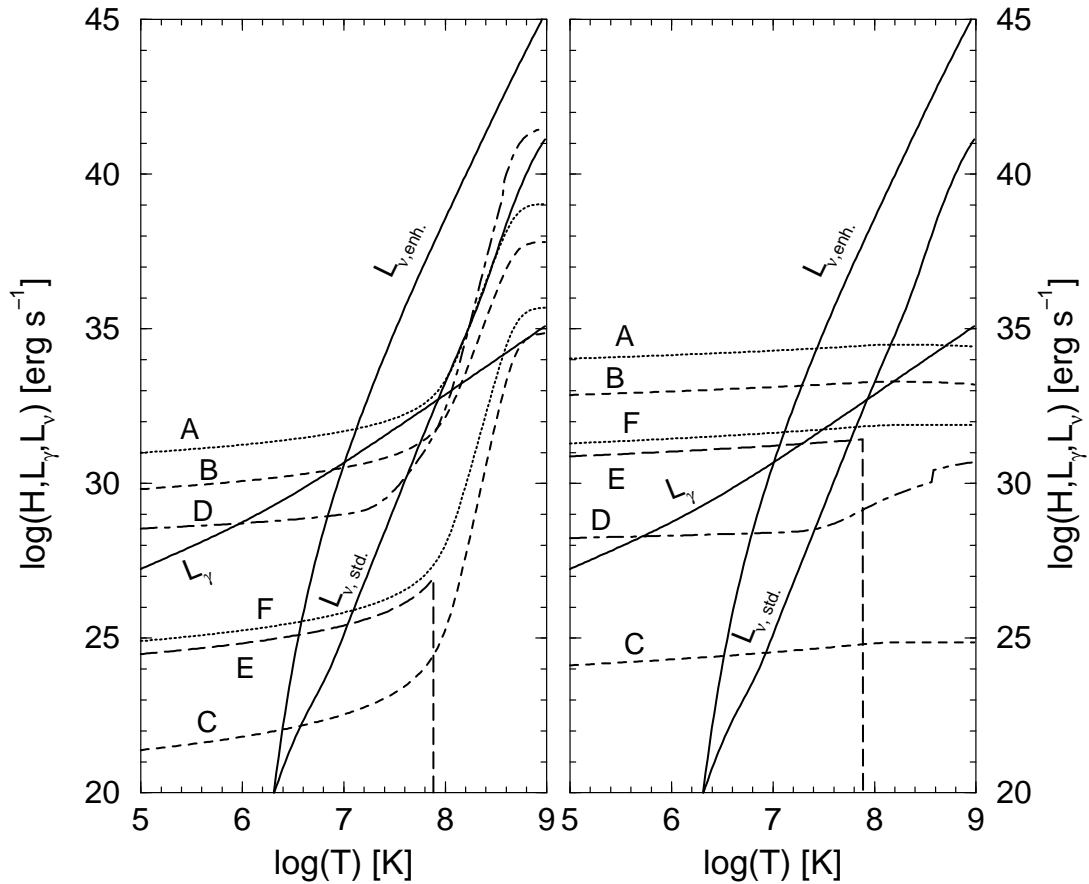}}}
\caption{Heating rate $H$ as functions of internal temperature for A: 
EB-pinning, B: PVB-pinning, C: corotating crust, D: corotating core,
E: crust cracking, F: chemical heating of the crust. The left panel
corresponds to $K=10^{-15}$~s, the right one to $K=10^{-22}$~s. The
photon luminosity, as well as the neutrino luminosity of a standard
cooling and an enhanced cooling model are also shown.}  \label{fig:lum}
\end{figure*}

In Fig. \ref{fig:lum} the heating rates
\begin{equation}
  H = \int 4\pi r^2 \e^{\Lambda+2\phi} h \df r
\end{equation}
discussed in Sect. \ref{sec:vortex} and \ref{sec:nonequilibrium} are
compared with the total photon and neutrino luminosities $L_\gamma$
and $L_\nu$, respectively, where an isothermal configuration,
i.e. $T\e^\phi =\mathrm{const.}$, is assumed. Since the heating rates
depend not only on the temperature but also on the time, we use the
standard thermal evolution scenario without heating to link the star's
temperature to its age.

We use two values of the rotational parameter $K$ (see
Eq. \eqref{eq:rot.evol}). About the first value, $K=10^{-15}$~s (left
panel in Fig. \ref{fig:lum}), the observed values of most of the
pulsars scatter, particularly of the five pulsars in class A. However
there are four pulsars (see Table \ref{tab:observation}) which are
rather old, $\tau >6\times 10^7$~yr, and rotate very fast,
$P<6$~ms. These pulsars are supposed to be \emph{recycled} by
accretion of matter from a companion star (see, e.g.,
\cite{Lorimer96a}). Since their rotational parameter $K\sim10^{-22}$~s
is quite different from the one of all the other pulsars, we shall
study them separately (see Sect. \ref{sec:res.recycled} and right
panel of Fig. \ref{fig:lum}).

For the value $K=10^{-15}$~s the rate of only three heating mechanism,
EB- and PVB-pinning and corotating core, exceed both, the photon and
the neutrino luminosities. We expect (and obtain) therefore only for
these three heating mechanism a considerable effect (see
Sect. \ref{sec:res}). All these heating mechanism affect both the late
and the early cooling. This is in contrast to the case, where
$K=10^{-22}$~s is employed. Since the time derivative of the angular
velocity $\dot\Omega$ is small in the latter case, only the late
cooling is affected.  Besides the three heating mechanisms quoted
above, the other heating mechanisms result in a heating rate which is
larger than the photon luminosity, except the heating from a
corotating crust. The rate of chemical heating of the core depends
also on the deviation from chemical equilibrium and can be compared
with the other processes only in detailed cooling simulations. As we
will see in Sect. \ref{sec:res.recycled}, it considerably affects the
late cooling for $K=10^{-22}$~s.

\section{Observed data} \label{sec:observations}

Only a few of the known pulsars have been detected by the X-ray
observatories Einstein, EXOSAT, ROSAT, and ASCA during the last two
decades. We use a sample of 27 pulsars to compare the theoretical
cooling curves with the observed data (see
Table \ref{tab:observation}). Both the timing characteristics
(i.e. $\Omega$ and $\dot\Omega$) and at least an upper limit on the
effective surface temperature $T_{\rm eff}^\infty$ as measured at
infinity are known for these pulsars. Table \ref{tab:observation}
summarises these data. The effective surface temperatures are
specified together with their $2\sigma$ error range.

\begin{table*} \centering
\caption[]{Sample of observed data \label{tab:observation}}
\small
\renewcommand{\baselinestretch}{1.0}
\begin{tabular}{lccccccll}
\hline
Pulsar            & $P$     & $\dot P$ & $\log(\tau)$ & $\log(K)$     
       & $\log(T_{\rm eff}^\infty)$    & Category & Reference \\
                  & [ms]    & [$10^{-15}{\rm\,ss}^{-1}$] & [yr] & [s] 
       & [K]                         &       & \\
\hline
0531+21           &   33.40 &   420.96 & 2.97\dag & -13.9
       & $6.18^{+0.19}_{-0.06}$ & B    & 1 \\
(Crab) &&&&&&& \\
1509-58           &  150.23 &  1540.19 & 3.19 & -12.6      
       & $6.11\pm 0.10$         & B    & 2 \\
0540-69           &   50.37 &   479.06 & 3.22 & -13.6
       & $6.77^{+0.03}_{-0.04}$ & B    & 3,4 \\
0002+62           &  241.81 &          & $\sim 4$\dag & $\sim -13$      
       & $6.20^{+0.07}_{-0.27}$ & A,bb & 5 \\
0833-45           &   89.29 &   124.68 & $4.3\pm 0.3$\dag & -14.0
       & $6.24\pm0.03$                 & A,bb & 6 \\
(Vela) &&&&&    $5.88\pm 0.09$         & A,mH & 7 \\
1706-44           &  102.45 &    93.04 & 4.24 & -14.0
       & $6.03^{+0.06}_{-0.08}$ & B    & 8 \\
1823-13           &  101.45 &    74.95 & 4.33 & -14.1
       & $6.01\pm 0.02$         & C    & 9 \\
2334+61           &  495.24 &   191.91 & 4.61 & -13.0
       & $5.92^{+0.15}_{-0.09}$ & C    & 10 \\
1916+14           & 1181    &   211.8  & 4.95 & -12.6
       & $5.93$                 & D    & 11 \\
1951+32           &   39.53 &     5.85 & 5.03 & -15.7
       & $6.14^{+0.03}_{-0.05}$ & B    & 12 \\
0656+14           &  384.87 &    55.03 & 5.05 & -13.7
       & $5.98\pm0.05$ 		& A,bb & 13 \\
&&&&&    $5.72^{+0.06}_{-0.05}$ & A,mH & 14 \\
0740-28           &  167    &    16.8  & 5.20 & -14.6
       & $5.93$                 & D    & 11 \\
1822-09           &  769    &    52.39 & 5.37 & -13.4
       & $5.78$                 & D    & 11 \\
0114+58           &  101    &     5.84 & 5.44 & -15.2
       & $5.98\pm 0.03$         & C    & 11 \\
1259-63           &   47.76 &     2.27 & 5.52 & -16.0
       & $5.88$                 & C    & 15 \\
0630+18           &  237.09 &    10.97 & 5.53 & -14.6
       & $5.76^{+0.04}_{-0.08}$ & A,bb & 16 \\
(Geminga) &&&&&    $5.42^{+0.12}_{-0.04}$ & A,mH & 17 \\
1055-52           &  197.10 &     5.83 & 5.73 & -14.9
       & $5.90^{+0.06}_{-0.12}$ & A,bb & 18 \\
0355+54           &  156.38 &     4.39 & 5.75 & -15.2
       & $5.98 \pm 0.04$        & C    & 19 \\
0538+28           &  143.15 &     3.66 & 5.79 & -15.3
       & $5.83$                 & C    & 20 \\
1929+10           &  226.51 &     1.16 & 6.49 & -15.6
       & $5.52$                 & B    & 21 \\
1642-03           &  388    &     1.77 & 6.54 & -15.2
       & $6.01\pm 0.03$         & C    & 11 \\
0950+08           &  253.06 &     0.23 & 7.24 & -16.3
       & $4.93^{+0.07}_{-0.05}$ & B    & 22 \\
0031-07           &  943    &     0.40 & 7.56 & -15.4
       & $5.57$                 & D    & 11 \\
0751+18           &    3.47 & $7.9\times 10^{-4}$ & 7.83 & -20.6
       & $5.66$                 & C    & 23 \\
0218+42           &    2.32 & $8.0\times 10^{-5}$ & 8.66 & -21.7
       & $5.78$                 & C    & 24 \\
1957+20           &    1.60 & $1.7\times 10^{-5}$ & 9.18 & -22.6
       & $5.53$                 & C    & 25,26 \\
0437-47           &    5.75 & $3.8\times 10^{-5}$ & 9.20 & -21.7
       & $5.94\pm 0.03$         & B    & 27 \\
\hline
\end{tabular}
\comments{The entries are: rotation period $P$, spin-down age
$\tau=P/2\dot P$, $K=P\dot P$ (see Eq. \eqref{eq:rot.evol}), effective
surface temperature as measured at infinity $T_{\rm eff}^\infty$. The
four categories A to D are explained in the text. bb and mH refer to
blackbody and magnetic hydrogen atmosphere fits, respectively. \dag:
estimated true age instead of spin-down age (see text). References: 1:
\cite{Becker95a}, 2: \cite{Seward83a}, 3: \cite{Finley93a}, 
4: \cite{Boyd95a}, 5: \cite{Hailey95a}, Fig. 2 with lower limit on
$N_{\rm H}$, 6: \cite{Oegelman95a}, Table III, 7: \cite{Page96a},
Fig. 1, 8: \cite{Becker95b}, 9: \cite{Finley93b}, 10:
\cite{Becker96a}, 11: \cite{Slane95a}, 12: \cite{SafiHarb95b}, 13:
\cite{Possenti96a}, 14: \cite{Anderson93}, 15: \cite{Becker96c}, 
16: \cite{Halpern97a}, PSPC+SIS0, 17: \cite{Meyer94}, Fig. 2a with
$B_{12}=1.18$, 18: \cite{Greiveldinger96a}, Table 2, 19:
\cite{Slane94a}, 20: \cite{Sun96a}, 21: \cite{Yancopoulos94a}, 
22: \cite{Manning94a}, 23: \cite{Becker96d}, 24: \cite{Verbunt96a},
25: \cite{Kulkarni92a}, 26: \cite{Fruchter92a}, 27: \cite{Zavlin97a}.}
\end{table*}

The ages of all pulsars except PSRs 0531+21 (Crab), 0833-45 (Vela) and
0002+62 are estimated via their spin-down age, $\tau=P/2\dot P$. This
relation requires that both the moment of inertia and the magnetic
surface field are constant, and that the braking index $n$ is equal to
its canonical value of 3, which corresponds to a spin-down by emission
of pure magnetic dipole radiation (see Sect.\ \ref{sec:spin.normal}).
Of course, the spin-down age is a crude approximation to the true age
of a neutron star, which can deviate from this value by a factor as
large as $\sim 3$, as discussed before.

The situation is different for the three pulsars quoted above, where
ages are known from different sources, that is: the age of the Crab
pulsar is known from the chronicles, the age of Vela was recently
determined via the proper motion of the pulsar with respect to the
supernova remnant by Aschenbach et al. (1995)\nocite{Aschenbach95a},
and the approximate age of PSR 0002+62 is given by an estimate of the
age of the related supernova remnant G 117.7+06 (\cite{Hailey95a}).

The information obtained from the X-ray observations is not always
sufficient to extract the effective surface temperature of the
corresponding neutron star. The sample is therefore divided into four
categories labelled A through D (\cite{Oegelman95a,Schaab95a}):
\begin{description}
\item[Category D:] 
The four pulsars have not been detected in the soft X-ray range so
far.  By considering the instrumental sensitivity an upper limit for
the surface temperature could be set.  These pulsars are marked with
white triangles in the figures below.
\item[Category C:]
The detections of ten pulsars contain too few photons for spectral fits. The 
surface temperatures were obtained by using the totally detected 
photon flux.  These pulsars are marked with black triangles.
\item[Category B:]
The spectra of eight pulsars, including the Crab pulsar 0531+21, can 
only be fitted by a power--law--type spectrum or by a blackbody spectrum with 
very high effective temperatures and effective areas much smaller than a 
neutron star surface. Presumably, their X-ray 
emission is dominated by magnetospheric 
emission. Therefore, the temperatures, determined from the spectral fits, are 
probably higher than the actual surface temperatures. 
Pulsars of this type are marked with arrows.
\item[Category A:]
Finally, there are five pulsars, 0833-45 (Vela), 0656+14, 0002+62, 
0630+18 (Geminga), and 1055-52, which allow two--component spectral fits. The 
softer blackbody component is believed to correspond to the actual surface 
emission of the neutron star, while the harder blackbody (or power--law) 
component may be due to magnetospheric emission.  These pulsars are marked 
with errorbars.
\end{description} 

The obtained effective surface temperature of pulsars of category A
depend crucially on whether or not a hydrogen atmosphere is
assumed. PSR 0833-45 (Vela) is a specific example (cf. Table
\ref{tab:observation}). Generally a spectral fit with a magnetic
hydrogen atmosphere yields a substantially lower effective surface
temperature than a blackbody fit. These models of hydrogen atmospheres
however do not yet take into account the presence of neutral atoms
(\cite{Potekhin96c}). Their effect should yield a spectrum that is
more similar to the blackbody spectrum, as indicated by the
preliminary estimates by Shibanov et al. (1993)\nocite{Shibanov93a}.
The atmospheric composition of a specific pulsar could be determined
by considering multiwavelength observations in the near future
(\cite{Pavlov96a}). We shall use the effective surface temperatures
obtained by fitting the spectra with both the blackbody and the
magnetic hydrogen model atmosphere for the three pulsars of category
A, for which both fits were accomplished.

\section{Results and discussion} \label{sec:res}

\subsection{Standard cooling}

\begin{figure}
\resizebox{\hsize}{!}{\includegraphics{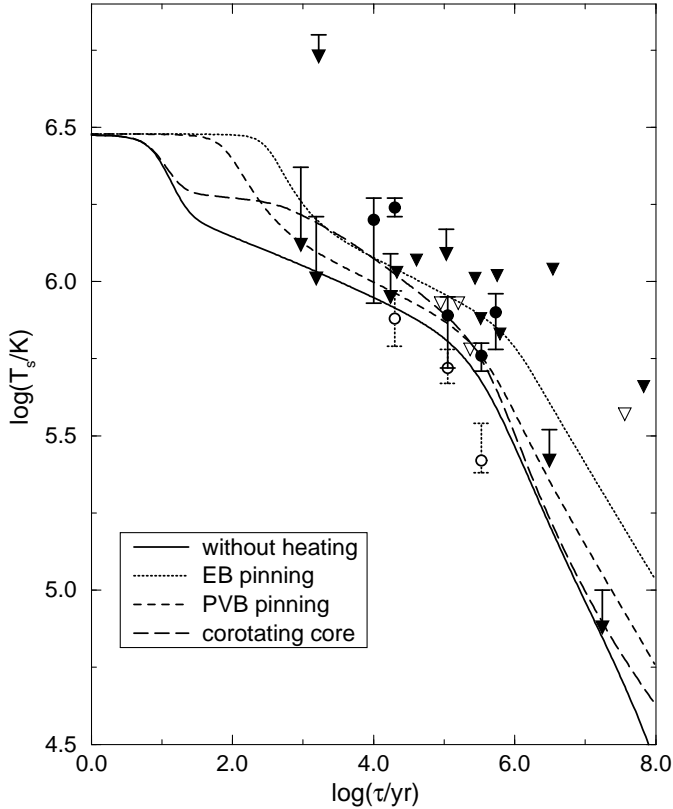}}
  \caption[]{Standard cooling with and without internal
  heating. Parameters are: \UVU\ equation of state, $M=1.4M_\odot$,
  and $K=10^{-15}$~s. See Table \ref{tab:observation} for the
  observed data. \label{fig:cool1}}
\end{figure} 
Figure \ref{fig:cool1} shows the surface temperature as measured by a
distant observer as a function of the star's age for the slowly
cooling neutron star models with and without internal heating. These
models are based on the non-relativistic equation of state \UVU\ which
account only for chemically equilibrated nucleons and leptons
(cf. Sect. \ref{sec:inp.eos}). We choose the canonical value
$M=1.4M_\odot$ for the gravitational mass of a neutron star.  Note
that the relatively low surface temperature is caused by the inclusion
of the superfluid pair breaking and formation process
(\cite{Flowers76b,Voskresenskii87a,Schaab95b}), which was not taken
into account in the earlier investigations (e.g. \cite{Umeda94};
\cite{Page95}; \cite{Schaab95a}). The standard cooling model without
internal heating is consistent with most of the observations.  A
discrepancy occurs however for the effective surface temperature of
the Vela pulsar 0833-45 derived for a blackbody fit, which tends to
lie considerably above all cooling tracks.  The effective radius
obtained for the blackbody fit, i.e. $R_{\rm eff}\approx 6{\rm\, km}$,
is considerably smaller than the canonical value of $\sim 10 {\rm
km}$, generally obtained for standard equations of state. Therefore,
the existence of a magnetic hydrogen (or helium) atmosphere is more
reliable and in agreement with the standard cooling scenario without
internal heating.

The effect of heating of the crust due to thermal creep of pinned
vortices is shown by the dotted and by the short dashed curves in
Fig. \ref{fig:cool1} for the EB- and PVB-pinning model, respectively
(see Sect.  \ref{sec:pinning}). The heating rate depends on
$\dot\Omega$ (see Eq.  [\ref{eq:pinn.heatrate}]) and thus on the inverse
of the magnetic dipole moment. The employed value $K=10^{-15}$\,s (see Eq.
[\ref{eq:rot.evol}]) is in the range $10^{-16}$--$10^{-13}$\,s
about which the observed values tend to scatter, with the exception of
the four millisecond pulsars which will be addressed in Sect.
\ref{sec:res.recycled}.  The effective surface temperature is
increased in both models in a similar way, though the effect of the
EB-pinning model is larger because of the larger pinning energy in the
inner crust. The heating of the crust can easily be recognised by the
increase of the thermal diffusion time, i.e. by the amount of time
needed for the cooling wave to reach the surface
(\cite{Lattimer94a}). After this time the surface temperature drops
significantly. It is increased by a factor of $\sim 50$ ($\sim 10$) in
the case of EB- (PVB-)pinning. This characteristic feature will become
very interesting if the thermal spectrum of a young pulsar
$\tau\lesssim 100{\rm\, yr}$ should be detected.

For $\tau\gtrsim 10^6{\rm\, yr}$ the slope of the cooling track is
increased by the heating of the crust. The observed upper limit of the
effective surface temperature of PSR 0950+08 seems to rule out the
EB-pinning model. Although consideration of a magnetic photosphere
would decrease the slope again, compensation of the heating effect
would require an unphysically strong magnetic field
(\cite{VanRiper88}). For middle aged pulsars, $10^3{\rm\,
yr}<\tau<10^6{\rm\, yr}$, heating due the PVB-pinning model has only a
small effect compared to the other uncertainties of the input
parameters.

The other three heating processes possible in the crust, driven by
drag forces, crust cracking, and chemical heating have no visible
effect on the cooling, since the heating rates are too small in
comparison with the neutrino and photon emission rates (for the value
of $K$ used here).  In the case of crust cracking our results are in
contrast to the ones reported by Cheng et al. (1992)\nocite{Cheng92a},
where a significant effect especially during the late cooling epoch
has been obtained. This deviation can be traced back to different
expressions used for the heating rates (see Sect.
\ref{sec:noneq.cracking})

Electron-vortex scattering in the interior of the star affects the
stellar cooling at times in the range $10^1<\tau<10^6{\rm\, yr}$ (see
Fig. \ref{fig:cool1}).  During this period the effect is comparable to
the effect of EB-pinning in the crust.  Heating due to vortex decay at
the crust-core boundary yields an increase of the surface temperature
only for rather old pulsars ($\tau>10^7{\rm\, yr}$). It can not be
constrained by any present pulsar observation, however.

\subsection{Enhanced cooling}

\begin{figure}
\resizebox{\hsize}{!}{\includegraphics{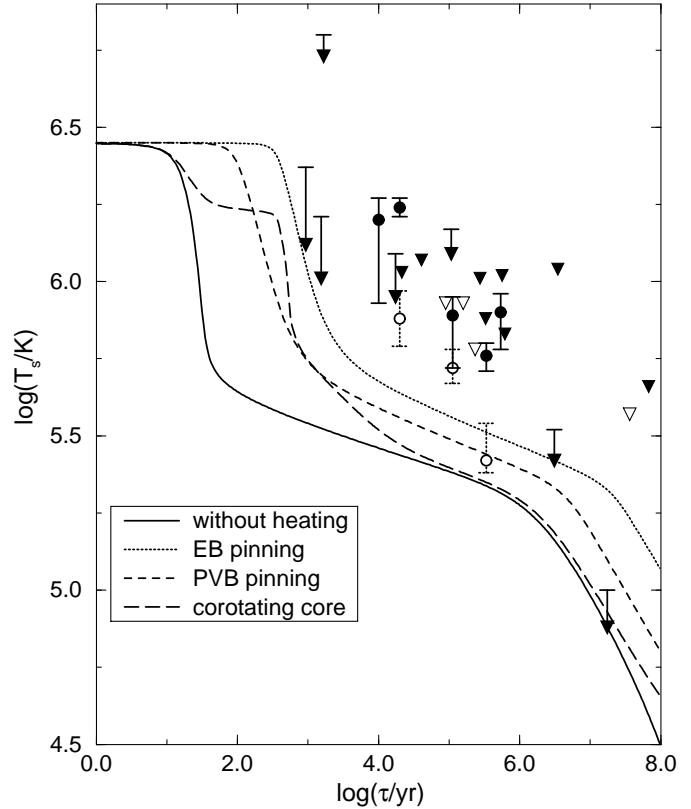}}
  \caption[]{Enhanced cooling of neutron star models constructed for
  	the RHF8 equation of state. Other parameters are as in
  	Fig. \ref{fig:cool1}. \label{fig:cool4}}
\end{figure} 
Figure \ref{fig:cool4} shows the cooling behaviour of enhanced cooling
models which are based on the RHF8 equation of state. The surface
temperature drops by a factor of $\sim 6$ when the cooling wave has
reached the surface. The fast direct Urca processes are only
suppressed below the critical temperature of neutron and lambda
pairing, $T_{\rm c}\sim2.0\times 10^9$~K and $T_{\rm c}\sim 1.6\times
10^9$~K, respectively.

The effect of heating on the cooling behaviour is similar as in the
case of standard cooling. The EB-pinning model shows again the largest
effect on the surface temperature. 

With the only exception of PSR 0630+18 (Geminga), the effective
surface temperatures of the pulsars of category A are clearly not
consistent with the enhanced cooling models, no matter whether one
includes heating processes or not. This discrepancy can be reduced
either by enhancing the gap energies (\cite{Page95,Schaab95a}) or by
assuming an accreted envelop.

\subsection{Accretion} \label{sec:res.accr}

\begin{figure}
\resizebox{\hsize}{!}{\includegraphics{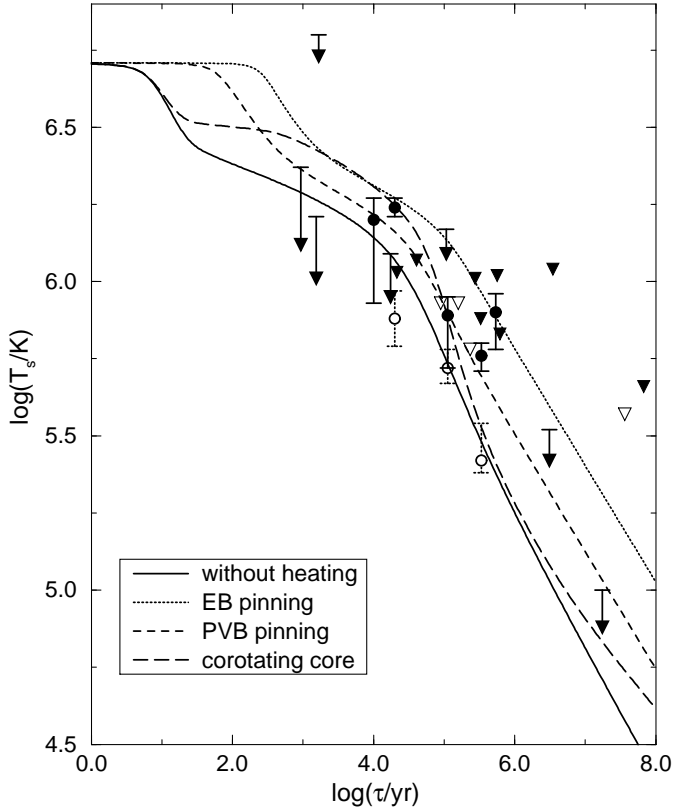}}
  \caption[]{Standard cooling with fully accreted envelope. Other
  	parameters are as in
  	Fig. \ref{fig:cool1}. \label{fig:cool5}}
\end{figure} 
\begin{figure}
\resizebox{\hsize}{!}{\includegraphics{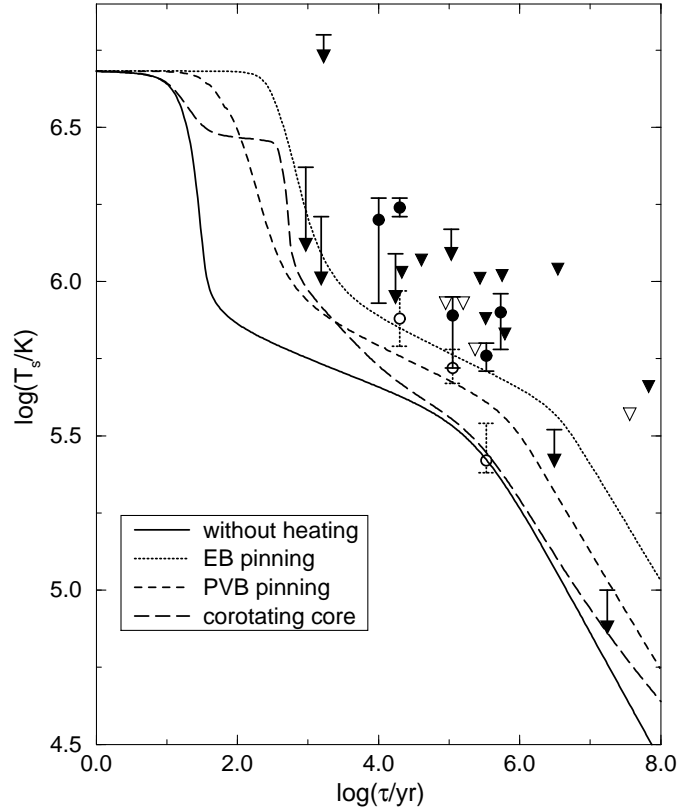}}
  \caption[]{Enhanced cooling with fully accreted envelope. Other
  	parameters are as in
  	Fig. \ref{fig:cool4}. \label{fig:cool7}}
\end{figure} 
In Figs. \ref{fig:cool5} and \ref{fig:cool7}, we study the effect of a
fully accreted envelope, which means that the accreted mass exceeds
$\sim 10^{-7}M_\odot$ (see \cite{Potekhin96c} and
Sect. \ref{sec:inp.photo}). Due to the envelope consisting of light
atoms such as hydrogen, helium and carbon, the temperature gradient of
the photosphere is substantially reduced. This yields a high effective
surface temperature in the neutrino cooling era, during which the
energy loss rate depends on the internal temperature; whereas it
yields small surface temperatures in the photon cooling era where the
loss rate depends on the surface temperature itself.

As it was already noted, the enhanced cooling models are now in better
agreement with the data of observation class A (see
Fig. \ref{fig:cool7}). This is especially true for the surface
temperatures obtained within the magnetic hydrogen atmosphere
model. The enhanced cooling model with heating due PVB-pinning fits
almost perfectly these surface temperatures.

\subsection{Millisecond pulsars} \label{sec:res.recycled}

\begin{figure}
\resizebox{\hsize}{!}{\includegraphics{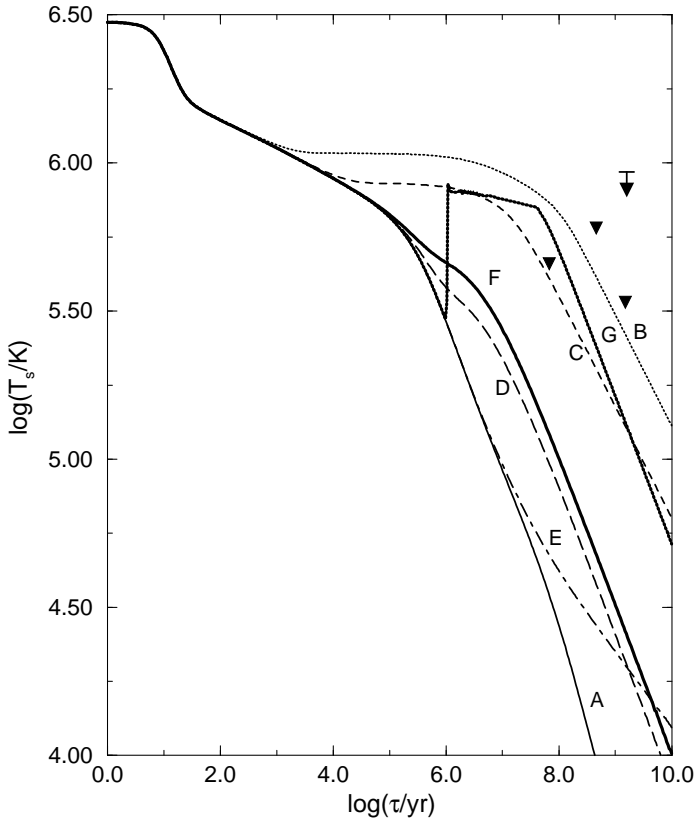}}
  \caption[]{Same as Fig. \ref{fig:cool1} but with various heating
  mechanism and $K=10^{-22}$~s. The curves correspond to the models
  without internal heating (labeld with A), with EB-pinning (B),
  PVB-pinning (C), crust cracking (D), corotating core (E), chemical
  heating of the crust (F), and chemical heating of the core (G),
  respectively. \label{fig:cool6}}
\end{figure}

So far we have investigated cooling scenarios associated with a large
rotational parameter, i.e. $K=10^{-15}$~s (see
Eq. [\ref{eq:rot.evol}]). This value is consistent with the bulk
fraction of observed pulsars (cf. Table \ref{tab:observation}). The
four oldest pulsars, PSR 0751+18, 0218+42, 1957+20, and 0437-47,
however have much smaller values around $K\sim 10^{-22}$~s. As it was
shown in Sect. \ref{sec:inp.comp}, this causes considerably different
heating rates. In particular, the mechanism of chemical heating and
crust cracking then becomes efficient enough to noticeably influence
the thermal evolution. In Fig. \ref{fig:cool6} we show the results for
the various heating mechanism in comparison with models without
heating. Since the late cooling behaviour (for $\tau\gtrsim 10^7$~yr)
does not depend on the cooling scenario, we consider here only the
standard cooling scenario. The model with chemical heating of the core
shows a sudden rise of the surface temperature at $\tau\sim
10^6$~yr. This is caused by the release of chemical energy through the
$\beta$-decay of neutrons. This process is suppressed by superfluidity
if the deviation from chemical equilibrium is smaller than the gap
energy (s. Sect. \ref{sec:chem} and \cite{Reisenegger96a}).

All cooling tracks in Fig. \ref{fig:cool6} are consistent with the
upper temperature limits of the three oldest millisecond pulsars. Only
the observation of PSR 0751+18 seems to exclude the EB-pinning
model. The situation changes however if one assumes that the actual
temperatures are not too far from  the upper limits. Then
only the PVB- and EB-pinning models, as well as the chemical heating
of the superfluid core yield sufficiently high surface temperature. A
future temperature determination, which will set a firm lower
limit on the temperature for one of these
pulsars could therefore rule out at least the corotation and the crust
cracking models as the only heat sources in millisecond pulsars.

\subsection{Two dimensional simulations}

In this section we relax the simplification that the heating rates are
angle independent and extend our one-dimensional calculations of the
previous sections to the two-dimensional case.  The averaging of the
local heating rates over spherical shells assumes that the heat
conductivity is very high in transverse direction. This is indeed a
fairly good approximation, as we shall see next by comparing the
results of one dimensional calculation with those of fully
two-dimensional ones.  The two-dimensional code was developed recently
by Schaab \& Weigel (1998)\nocite{Schaab98c} (s.a. \cite{Schaab97a}).

\begin{figure}
\resizebox{\hsize}{!}{\includegraphics{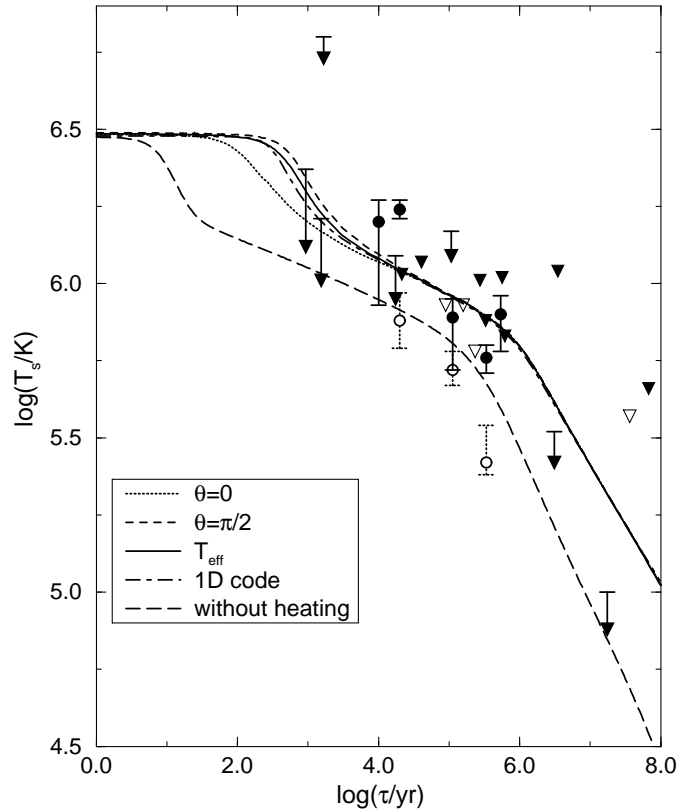}}
  \caption[]{Comparison of the one and two dimensional cooling
  simulations for the \UVU --model with inclusion of EB-pinning.
  \label{fig:cool8}}
\end{figure}

Fig. \ref{fig:cool8} shows the obtained cooling tracks for standard
cooling computed for the \UVU --model with inclusion of the EB-pinning
model. Heating leads to a large temperature difference (up to 50\,\%)
between the pole ($\theta=0$) and the equator ($\theta=\pi/2$) for a
pulsar whose age lies in the range
$50\lesssim\tau\lesssim10^4$~yr. Moreover the temperature drop at the
equator is delayed by the heating with respect to the model without
heating and with respect to the drop of the polar temperature. This is
the result of the $h\propto\sin\,\theta$ dependence for the vortex
creep model.

The surface temperature obtained with the one dimensional code can be
compared with the effective temperature which is defined by
(\cite{Page95c})
\begin{equation}
  T_{\rm eff} = \left(\int_0^1 \df s ~ 2 s ~
    \left( \e^\nu T(\theta)\right)^4 \right)^{1/4},
\end{equation}
where $s=\sin\,\delta$.
The emission angle $\delta$ is related to the colatitude $\theta$ by
the equation
\begin{equation} \label{eq:geodes}
  \theta(\delta) = \int_0^{M/R}\df u \frac{\sin\,\delta}
    {\left(\left(1-\frac{2M}{R}\right)\left(\frac{M}{R}\right)^2
    -(1-2u)u^2\sin^2\,\delta\right)^{1/2}},
\end{equation}
where $M$ and $R$ are the neutron star's mass and radius,
respectively. In flat spacetime, equation \eqref{eq:geodes} simplifies
to $\delta =\theta$.

The effective temperature $T_{\rm eff}$ and the surface temperature
obtained with the one dimensional code are almost identical
(see Fig. \ref{fig:cool8}). This means that the luminosity obtained
with both codes are also identical. The results of the one dimensional
codes are thus acceptable for a comparison with the observed data
of categories B--D, where the upper bounds on the surface temperature are
derived from the respective luminosity. If one uses spectral
information as for the pulsars in category A, the observations should be
compared with the spectra derived from the surface temperature
distribution $T_{\rm s}(\theta)$ in the framework of two dimensional
simulations. Nevertheless, we expect that the conclusions remain
unchanged if we use the effective temperature also for a comparison
with the observed data of category A.

\section{Conclusions} \label{sec:disc}

In this work, we have carried out a comparative analysis of the impact
of different competing heating processes on the cooling behaviour of
neutron stars.  In general we find that the internal heating yields
significantly enhanced surface temperatures of middle aged and old
pulsars, as one would expect. However, the effectiveness of these
heating processes varies significantly from one process to another,
and depends sensitively on the value of the rotational parameter $K$.

We studied models with two different rotational parameters.  The first
value was chosen to amount $K\sim 10^{-15}$~s, which is supported by
pulsars observations.  For this $K$ value, the heating due to thermal
creep of pinned vortices and motion of proton vortices in the interior
of the star yield considerable enhancements of the surface
temperatures of middle aged pulsars with respect to the models which
ignore heating. This leads to closer agreement with the observed data
in the case of enhanced cooling, which is even more improved for the
case of the fully accreted envelopes (s.a. \cite{Page96b}).

The observed upper temperature limit for PSR 0950+08 seems to rule out
the strong pinning models, whereas the weaker pinning models are
consistent with the observations (the pinning models differ by the
value of the pinning energy in the inner crust).  The other heating
processes -- from chemical heating to dissipative motion of neutron
vortices in the crust to crust cracking -- have no observable effect
on the cooling. Our result for heating due to crust cracking is in
contrast to the one obtained by Cheng et al. (1992)\nocite{Cheng92a},
who find a large effect especially on the late cooling stages. The
difference arises from different expressions adopted for the heating
rates.

The four oldest pulsars of our sample rotate very fast with periods of
a few milliseconds. Hence the rotational parameter for millisecond
pulsars is smaller than that for the bulk fraction of the population
by 7 orders of magnitude ($K=10^{-22}$~s). For this value only the
late cooling stages are affected by heating. We find that the heating
due to thermal creep of pinned vortices and chemical heating of the
core has the largest impact on the surface temperatures of millisecond
pulsars.  Again, the strong pinning models leads to surface
temperatures which are larger than one of the upper limits deduced for
the millisecond pulsars.

As far as we know, the published cooling simulations of neutron stars
which account for internal heating were performed in one spatial
dimension.  Such a treatment is only possible if the angle dependent
heating rates are averaged over spherical shells. By comparing the
outcome of one-dimensional simulations with fully two-dimensional
simulations, based on a recently developed numerical code
(\cite{Schaab98c}), we could confirm the validity of this procedure.

The knowledge of the input parameters for cooling simulations such as
the composition, the superfluid gap energies, and the possible
existence of an accreted envelope, etc. is far form being complete.
The number of parameters involved in the cooling simulations increases
when the internal heating processes are taken into account.  In
principal, one could expect to derive some of the parameters
associated with these phenomena by means of comparing the theoretical
cooling models with the body of observed data.  This attempt needs to
be supplemented with the analysis of terrestical experiments, other
astrophysical observations, as well as future theoretical studies.
Nevertheless, the observed data already constrain the microphysical
input, as, for instance, for the scenaria of heating via vortex creep.

Tables containing detailed references to the ingredients used in the
simulations, the observational data, and the resulting cooling tracks
can be found on the Web:
http://www.physik.uni-muenchen.de\hspace{0pt}/sektion\hspace{0pt}/suessmann\hspace{0pt}/astro\hspace{0pt}/cool.

\section*{Acknowledgments}

We would like to thank H.-T. Janka for helpful discussions and the
anonymous referee for many valuable suggestions. Ch.~S. gratefully
acknowledges the Bavarian State for a fellowship.  A.~S. has been
supported through a research grant from the Max Kade Foundation.


\end{document}